# Static global monopoles in higher dimensional space time

R.Mukherjee and F.Rahaman

Department of Mathematics

Jadavpur University, Kolkata – 700 032, India

E-mail: farook_rahaman@yahoo.com

## Abstract:

We present an exact solution around global monopole resulting from the breaking of a global S0(3) symmetry in a five dimensional space time. We have shown that the global monopole in higher dimensional space time exerts gravitational force which is attractive in nature. It is also shown that the space around global monopole has a deficit solid angle. Finally, we study monopole in higher dimensional space time within the framework of Lyra geometry.

## Introduction:

The idea of higher dimensional theory was originated in Super String and Super Gravity theories with the other fundamental forces in nature. To find a theory which unifies gravity with the other forces in nature remains an open problem in quantum field theory even today. Developments in Super String theories have stimulated the study of physics in higher dimensional space times [1]. More over, solutions of Einstein field equations in higher dimensional space times are believed to be physical relevance possibly at the extremely early times before the Universe underwent the compactification transitions. As a result higher dimensional theory is receiving great attention both in cosmology and particle physics. In quantum field theory, when a symmetry has broken during the phase transitions, several topological defects will arise [2]. Global monopole ( a kind of topological defect which is formed when a global symmetry is broken ) is important objects both Particle Physicists and Cosmologists which predicted to exist in Grand Unified Theory. Using a suitable scalar field it was shown that the phase transitions on the early Universe can give rise to such objects which are nothing but the topological knots in the vacuum expectation value of the scalar field and most of their energy is concentrated in a small region near the monopole core.

In 1989, Barriola and Vilenkin (BV) [3] have shown an approximate solution of the Einstein equations for the metric out side a global monopole, resulting from a global S0(3) symmetry breaking. Banerjee et al [4] has extended the work of BV to higher dimension. Their space time has the topology of $R^1 \times S^1 \times S^2 \times S^1$. Their five dimensional monopole metric was not unique where as BV monopole metric was unique.

In this work, we would like to consider global monopole in higher dimensional space time with topology is $R^1 \times S^1 \times S^3$. The motivation of this work is to look forward whether the global monopole shows any significant properties due to consideration of the space time with topology $R^1 \times S^1 \times S^3$.

While attempting to unify gravitation and electromagnetism in a single space time geometry, Weyl [5] showed how one can introduce a vector field with an intrinsic geometrical significance. But this theory was not accepted as it was based on non-integrability of length transfer. Lyra [6] proposed a new modifications of Riemannian geometry by introducing a gauge function which removes the non-integrability condition of a vector under parallel transport.

In consecutive investigations Sen [7] and Sen and Dunn [8] proposed a new scalar tensor theory of gravitation and constructed an analog of Einstein field equations based on Lyra's geometry which in normal gauge may be written as

$$R_{ab} - \tfrac{1}{2} g_{ab} R + \tfrac{3}{2} *\varphi_a *\varphi_b - \tfrac{3}{4} g_{ab} *\varphi_c *\varphi^c = -8\pi G T_{ab} \qquad \ldots\ldots(1)$$

where $*\phi_a$ is the displacement vector and other symbols have their usual meaning as in Riemannian geometry.

According to Halford [9], the present theory predicts the same effects within observational limits, as far as the classical solar system tests are concerned, as well as tests based on the linearized form of field equations. Soleng [10] has pointed out that the constant displacement field in Lyra's geometry will either include a creation field and be equal to Hoyle's creation field cosmology or contain a special vacuum field which together with the gauge vector term may be considered as a cosmological term.

Subsequent investigations were done by several authors in scalar tensor theory and cosmology within the frame work of Lyra geometry [11]. Recently, Rahaman et al and other authors have studied some topological defects within the framework of Lyra geometry[12].

In the present work, we also derive the solutions for the higher dimensional space time metric out side a global monopole within the framework of Lyra geometry in normal gauge i.e. displacement vector $*\phi_a = (\beta, 0, 0, 0, 0)$, where $\beta$ is a constant.

In section 3, we have studied higher dimensional global monopole in Lyra geometry.
Motion of the test particle in the gravitational field of higher dimensional global monopoles are discussed in section 4. The paper ends with a short discussion in section 5.

## 2. Global monopole in general relativity:

In this section we closely follow the formalism of BV and take the Lagrangian as

$$L = \tfrac{1}{2} \partial_\mu \Phi^i \partial^\mu \Phi^i - \tfrac{1}{4} \lambda (\Phi^i \Phi^i - \eta^2)^2 \qquad \ldots(2)$$

where $\Phi^i$ is a multiplet of scalar fields, i = 1,2,3,4 (where $\eta$ is the energy scale of symmetry breaking and $\lambda$ is a constant).

The field configuration describing a monopole is taken as

$$\Phi^i = \eta\, f(r)\, (x^i / r) \qquad \ldots(3)$$

where $x^i x^i = r^2$.

[ Actually $(x^i / r) \equiv n^i$ is a unit vector $(n^i n^i = 1)$ in dimensional Euclidean space with components $n^4 = \cos\psi$, $n^3 = \sin\psi \cos\varphi$, $n^2 = \sin\psi \sin\varphi \cos\theta$, $n^2 = \sin\psi \sin\varphi \sin\theta$ ]

The metric ansatz describing a monopole can be taken as

$$ds^2 = -A(r)\, dt^2 + B(r)\, dr^2 + r^2\, (d\theta^2 + \sin^2\theta\, d\varphi^2 + \sin^2\theta \sin^2\varphi\, d\psi^2) \qquad \ldots(4)$$

Using the Lagrangian (2) and metric (4) the components of energy momentum tensors can be written via [4]

$$T_{ab} = 2(\partial L / \partial g^{ab}) - L\, g_{ab} \qquad \ldots(5)$$

as follows:

$$T_t^t = \eta^2 [(f^1)^2 / 2B] + (3/2)\eta^2 (f^2 / r^2) + \tfrac{1}{4}\lambda (\eta^2 f^2 - \eta^2)^2 \qquad \ldots(6)$$

$$T_r^r = -\eta^2 [(f^1)^2 / 2B] + (3/2)\eta^2 (f^2 / r^2) + \tfrac{1}{4}\lambda (\eta^2 f^2 - \eta^2)^2 \qquad \ldots(7)$$

$$T_\theta^\theta = T_\varphi^\varphi = T_\psi^\psi = \eta^2 [(f^1)^2 / 2B] + \tfrac{1}{4}\lambda (\eta^2 f^2 - \eta^2)^2 + \tfrac{1}{2}\eta^2 (f^2 / r^2) \qquad \ldots(8)$$

( prime denotes the differentiation w.r.t. 'r' )

It can be shown that in flat space the monopole core has a size $\delta \sim \sqrt{\lambda}\, \eta^{-1}$ and mass, $M_{core} \sim \lambda^{-1/2}\, \eta$. Thus if $\eta << m_p$ where $m_p$ is the plank mass, it is evident that we can still apply the flat space approximation of $\delta$ and $M_{core}$.
This follows from the fact that in this case the gravity would not much influence on monopole structure.

Banerjee et al assumed that $f = 1$ out side the monopole core [4].
With this result the energy stress tensors assume the following form

$$T_t^t = T_r^r = 3(\eta^2 / 2r^2)\,;\qquad T_\theta^\theta = T_\varphi^\varphi = T_\psi^\psi = \tfrac{1}{2}(\eta^2 / r^2) \qquad \ldots(9)$$

$$3\,[B^1 / (2rB^2)] - [3/(Br^2)] + (3/r^2) = 12\pi G(\eta^2 / r^2) \qquad \ldots(10)$$

$$-[3/(Br^2)] + (3/r^2) - 3[A^1/(2rAB)] = 12\pi G(\eta^2 / r^2) \qquad \ldots(11)$$

$$[A^1 B^1 / 4AB] - [A^{11}/2AB] + [(A^1)^2/(4BA^2)] + [B^1/(rB^2)] -$$
$$[A^1/ABr] - [1/(Br^2)] + (1/r^2) = \tfrac{1}{2}(\eta^2 / r^2) \qquad \ldots(12)$$

Equation (10), we get

$$Z^{1} + 2(Z/r) = -2[(1 - 4\pi G\eta^{2})/r] \quad \ldots\ldots(13)$$

where $Z = -(1/B)$

Solving this equation, we get

$$B = [1 - 4\pi G\eta^{2} - (C/r^{2})]^{-1} \quad \ldots(14)$$

where C is an integration constant.

Now subtracting eq.(11) from eq.(12), we get

$$3[B^{1}/(2rB^{2})] + 3[A^{1}/(2rAB)] = 0 \quad \ldots(15)$$

This implies
$$AB = 1 \quad \ldots(16)$$

( without any loss of generality we can take the integration constant to be unity .)

Thus the solution is
$$A = B^{-1} = [1 - 4\pi G\eta^{2} - (C/r^{2})] \quad \ldots\ldots(17)$$

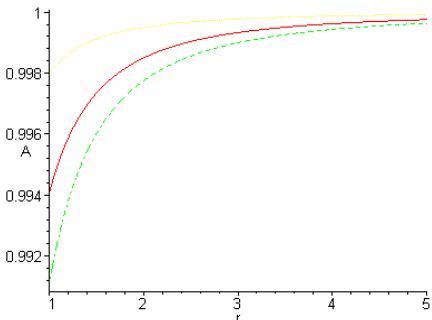

Diagram of the metric coefficient $g_{tt}$ for $8\pi G\eta^{2} = 10^{-6}$ and different values of C.
( C = .002, for yellow line; C = .006, for red line, C = .009, for green line )

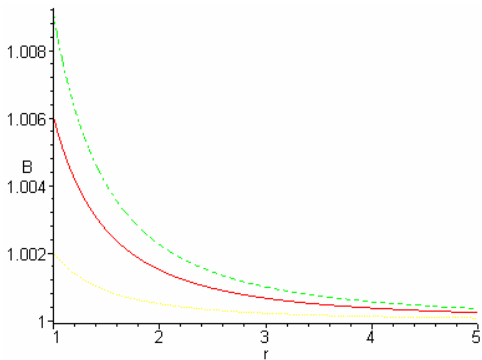

Diagram of the metric coefficient $g_{rr}$ for $8\pi G\eta^{2} = 10^{-6}$ and different values of C.
( C = .002, for yellow line; C = .006, for red line, C = .009, for green line )

It is of some interest to calculate bending of light in the above field in the plane $\theta = \frac{1}{2}\pi$.

The equation for the light track in the $\psi$ = constant hyper surfaces is

$$k^2 - h^2(dU/d\varphi)^2 - h^2 U^2 (1 - 4\pi G\eta^2 - CU^2) = 0 \qquad ....(18)$$

The constants $k$, $h$ are defined by

$$A(dt/dp) = k \text{ and } r^2 d\varphi/dp = h.$$

p being the affine parameter along the light path,

and

$$r = (1/U) \qquad .....(19)$$

From eq.(18), one get [ writing $\xi = (1 - 4\pi G\eta^2)^{\frac{1}{2}} \varphi$ ]

$$(d^2U/d\xi^2) + U[1 - 2CU^2(1 - 4\pi G\eta^2)^{-\frac{1}{2}}] = 0 \qquad ......(20)$$

If the light ray does not penetrate in to the monopole core, the last term is small and one

may write the above equation in the form

$$(d^2U/d\xi^2) + U = 2CU^3 \qquad .....(21)$$

The approximate solution of this equation is

$$U = U_0 \cos(\alpha\xi) + U_1 \cos(3\alpha\xi) \qquad ......(22)$$

with $U_1 = -C(U_0^2/16) << U_0$ and $\alpha^2 = 1 - 3(CU_0^2/2)$.

For $U = 0$,

one gets, $\alpha\xi = \pm \frac{1}{2}\pi$

or

$$\varphi = \pm \frac{1}{2}\pi (1 - 4\pi G\eta^2)^{-\frac{1}{2}} [1 - 3C(U_0^2/2)]^{-\frac{1}{2}} \qquad ......(23)$$

And bending comes out as

$$\pi[2\pi G\eta^2 + \frac{3}{4}CU_0^2] \qquad ......(24)$$

## 3. Global monopole in Lyra geometry:

In this section, we shall consider higher dimensional global monopole in Lyra geometry. We have taken the same energy momentum tensors as before.

The field equation (1) for the metric (4), reduces to

$$3[B^1/(2rB^2)] - [3/Br^2] + (3/r^2) - \tfrac{3}{4}(\beta^2/A) = 12\pi G(\eta^2/r^2) \quad ....(25)$$

$$-[3/(Br^2)] + (3/r^2) - 3[A^1/(2rAB)] + \tfrac{3}{4}(\beta^2/A) = 12\pi G(\eta^2/r^2) \quad ...(26)$$

$$[A^1 B^1/4AB] - \tfrac{1}{2}[A^{11}/AB] + [(A^1)^2/(4BA^2)] + [B^1/(rB^2)] - [A^1/ABr] - [1/(Br^2)]$$
$$+ 1/r^2 + \tfrac{3}{4}(\beta^2/A) = 4\pi G(\eta^2/r^2) \quad ...(27)$$

Now subtracting eq.(26) from eq.(25), we get

$$(AB)^1 = \beta^2 r B^2 \quad .....(28)$$

Since $\beta \neq 0$, we never get the general relativity like solution. According to BV, a global monopole solution should have $f = 1$ as $r \to \infty$. The dependence of $\eta^2$ of the asymptotic expansion of f is very weak. It appears that the asymptotic behavior of the monopole solution is quite independence of the scale of symmetry break down up to values as large as the planck scale [$\eta^2 = (4\pi G)^{-1}$]. However, in order to confirm the existence of monopole solutions up to $\eta^2 = (4\pi G)^{-1}$, we have to obtain the values of A and B from the field equations.

Adding eq.(25) with eq.(26), we get,

$$[B^1/(rB^2)] - (1/B)[(A^1/rA) + (4/r^2)] = 4[(1 - 4\pi G\eta^2)/r^2] \quad ....(29)$$

Using $\eta^2 = (4\pi G)^{-1}$, we get from eq.(29)

$$(B^1/B) = (A^1/A) + (4/r) \quad .....(30)$$

This implies
$$B = B_0 A r^4 \quad ...(31)$$

( here, $B_0$ is an integration constant )

Using eq.(30), from eq.(28), we get

$$2(A^1/A) + (4/r) = \beta^2 B_0 r^5 \quad ...(32)$$

Solving eq.(32), we get
$$A = (A_0/r^2) \exp[(1/12)\beta^2 B_0 r^6] \quad ......(33)$$

( here, $A_0$ is an integration constant )

Thus the higher dimensional monopole in Lyra geometry takes the following form

$$ds^2 = -(A_0/r^2)\exp[(1/12)\beta^2 B_0 r^6]\,dt^2 + B_0 A_0 r^2 \exp[(1/12)\beta^2 B_0 r^6]\,dr^2 + r^2\,d\Omega_3^2$$

....(34)

From the metric itself, it is quite apparent that there is no singularity at a finite distant from the monopole core.

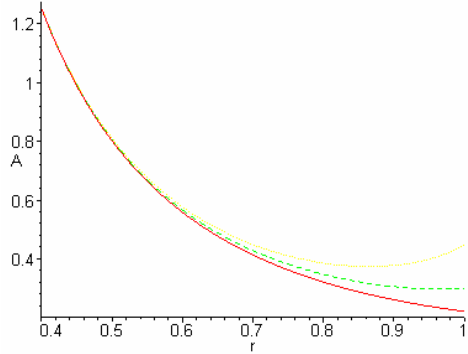

Diagram of the metric coefficient $g_{tt}$ for global monopole in Lyra geometry for different values of the displacement vector ( taking $A_0 = .2$, $B_0 = 1$ and $\beta^2 = 9.6$, for yellow line; $\beta^2 = 1.2$, for red line, $\beta^2 = 4.8$, for green line ).

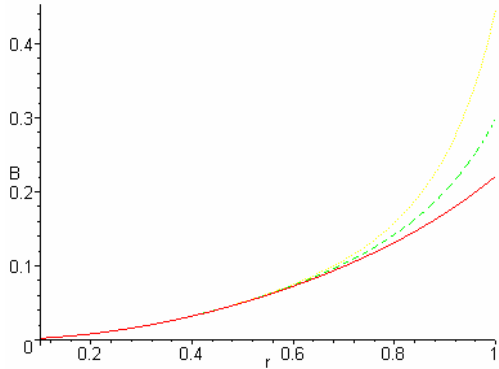

Diagram of the metric coefficient $g_{rr}$ for global monopole in Lyra geometry for different values of the displacement vector ( taking $A_0 = .2$, $B_0 = 1$ and $\beta^2 = 9.6$, for yellow line; $\beta^2 = 1.2$, for red line, $\beta^2 = 4.8$, for green line ).

## 4. Motion of a test particles:

Let us consider a relativistic particle of mass m moving in the gravitational field of the monopole described by eq. (4).
The Hamilton – Jacobi ( H-J) equation is [13]

$$-(1/A)(\partial S/\partial t)^2 + (1/B)(\partial S/\partial r)^2 + (1/r^2)[(\partial S/\partial x_1)^2 + (\partial S/\partial x_2)^2 + (\partial S/\partial x_3)^2] + m^2 = 0 \quad ....(35)$$

with $x_1$, $x_2$, $x_3$ are the co ordinates on the surface of the 3 – sphere.

Take the ansatz

$$S(t, r, x_1, x_2, x_3) = -E \cdot t + S_1(r) + p_1 \cdot x_1 + p_2 \cdot x_2 + p_3 \cdot x_3 \qquad \ldots(36)$$

as the solution to the H-J eq. (35).

Here the constant E is identified as the energy of the particle and $p_1, p_2, p_3$ are momentum of the particle along different axes on 3–sphere with $p = (p_1^2 + p_2^2 + p_3^2)^{1/2}$, as the resulting momentum of the particle.

Now substituting (36) in (35), we get

$$S_1(r) = \varepsilon \int [B\{(E^2/A) - (p^2/r^2) + m^2\}]^{1/2} \, dr \qquad (\text{where } \varepsilon = \pm 1) \qquad \ldots(37)$$

In H-J formalism, the path of the particle is characterized by [13]

$$(\partial S/\partial E) = \text{constant}, \ (\partial S/\partial p_i) = \text{constant} \ (i = 1,2,3) \qquad \ldots(38)$$

Thus we get (taking the constants to be zero without any loss of generality),

$$t = \varepsilon \int \{(\sqrt{B}) E / A\} [\{(E^2/A) - (p^2/r^2) + m^2\}]^{-1/2} \, dr \qquad \ldots(39)$$

$$x_i = \varepsilon \int \{(\sqrt{B}) p_i / r^2\} [\{(E^2/A) - (p^2/r^2) + m^2\}]^{-1/2} \, dr \qquad \ldots(40)$$

From (39), we get the radial velocity as

$$(dr/dt) = (A/E\sqrt{B}) [\{(E^2/A) - (p^2/r^2) + m^2\}]^{1/2} \qquad \ldots(41)$$

Now the turning points of the trajectory are given by $(dr/dt) = 0$ and as a consequence the potential curves are

$$(E/m) \equiv V = [A\{(p^2/m^2 r^2) + 1\}]^{1/2} \qquad \ldots(42)$$

We shall study the trajectory of the test particle for different situations:

**Case – I:** Global monopole in general relativity:

In this case the extremals of the potential curve are the solutions of the equation

$$r^2 [ (2p^2/m^2)( 1 - 4 \pi G\eta^2 ) - 2m^2C ] = 2Cp^2 ( 1 + m^{-2} ) \qquad \ldots(43)$$

We note that if $p^2 < C m^4( 1 - 4 \pi G\eta^2 )^{-1}$, then the above radical has no real extremals. Hence there is no window in the parameter space to produce bound states and particles can not be trapped by the monopole.
But this equation has at least one positive real root provided $p^2 > C m^4 ( 1 - 4 \pi G\eta^2 )^{-1}$.
So it is possible to have bound orbit for the test particle. Thus the gravitational field of the global monopole is shown to be attractive in nature but here we have to imposed some restriction relating symmetry breaking scale η and mass and momentum of the test particle.

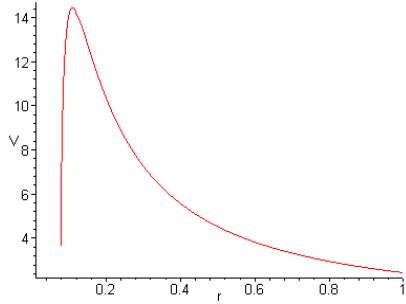

The diagram of the potential curve with respect to radial coordinate ( taking $(p^2/m^2) = 5$, $8\pi G\eta^2 = 10^{-6}$ and $C = .006$ ).

**Case – II:** Global monopole in Lyra geometry:

Here the extremals of the potential curve are the solutions of the equation

$$\beta^2 B_0 r^8 + (p^2 m^{-2}) \beta^2 B_0 r^6 - 4r^2 - 8p^2 m^{-2} = 0 \qquad \ldots(44)$$

This is an algebraic equation of even degree ( degree eight ) with negative last term. This equation has at least one real positive root. Thus bound orbit are possible in this situation. Hence the higher dimensional global monopole in Lyra geometry, always exert gravitational force which is attractive in nature.

## 5. Discussions:

At first, we state in brief, the nature of the BV's four dimensional monopole [3]. Solving the gravitational field equations and adopting some suitable scale changes, BV arrived at the metric

$$ds^2 = -dt^2 + dr^2 + (1 - 8\pi G\eta^2) r^2 d\Omega_2^2 \qquad \ldots(45)$$

From this they concluded:

(a) $g_{tt}^1 = 0$ i.e. the acceleration vector ($\mathbf{A^r}$) corresponding to the unit vector along time coordinate lines vanish, so the monopole exerts no gravitational force.
(b) The coefficient $(1 - 8\pi G\eta^2)$ of $r^2 d\Omega_2^2$ indicates a deficit solid angle.

For our higher dimensional monopole it is obvious that $\mathbf{A^r} \propto r^{-3}$. This shows that the gravitational force falls of as the inverse cube of the distances. We also see that the space time around our higher dimensional monopole has a deficit solid angle.

Banerjee et al higher dimensional monopole metric is not unique but our higher dimensional monopole metric is unique.

For large enough values of r, the solution (17) passes over to that given by BV.

A global monopole, however, is quite consistent in Lyra geometry and we obtain the exact solutions for the space time metric in some special case. The solutions represented by eq.(34) exhibits no singularity at a finite distance from the monopole core. This example is important as in general relativity all the solutions for a global monopole have a singularity for finite values of r.

Our higher dimensional monopole in general relativity exerts gravitational force which is attractive in nature provided some restriction to be imposed relating symmetry breaking scale η and mass and momentum of the test particle. This is quite similar to Banerjee et al monopole [4]. But higher dimensional global monopole in Lyra geometry always exert gravitational force, which is attractive in nature. Thus we see some important differences between higher dimensional global monopole in Lyra geometry with the classical result.

## Acknowledgments:

F.R is thankful to DST, Government of India for providing financial support.
We are also grateful to the anonymous referee for his valuable comments.